\documentclass{jfm}
\usepackage{graphicx}
\usepackage{epstopdf, epsfig}

\usepackage{color}

\shorttitle{Maximal heat transfer between two parallel plates}
\shortauthor{S. Motoki, G. Kawahara and M. Shimizu}

\title{Maximal heat transfer between two parallel plates}

\author{Shingo Motoki\aff{1}\corresp{\email{motoki@tes.me.es.osaka-u.ac.jp}}, Genta Kawahara\aff{1}\and Masaki Shimizu\aff{1}}
\affiliation{\aff{1}Graduate School of Engineering Science, Osaka University, 1-3 Machikaneyama, Toyonaka, Osaka 560--8531, Japan}

\begin{document}

\maketitle

\begin{abstract}
The divergence-free time-independent velocity vector field has been determined so as to maximise heat transfer between two parallel plates of a constant temperature difference under the constraint of fixed total enstrophy.
The present variational problem is the same as that first formulated by \cite{Hassanzadeh2014}; however, a search range of optimal states has been extended to a three-dimensional velocity field.
The scaling of the Nusselt number $Nu$ with the P\'eclet number $Pe$ (i.e., the square root of the non-dimensionalised enstrophy with thermal diffusion timescale), $Nu\sim Pe^{2/3}$, has been found in the three-dimensional optimal states, corresponding to the asymptotic scaling with the Rayleigh number $Ra$, $Nu\sim Ra^{1/2}$, in extremely-high-$Ra$ convective turbulence, and thus to the Taylor energy dissipation law in high-Reynolds-number turbulence.
At $Pe\sim10^{0}$, a two-dimensional array of large-scale convection rolls provides maximal heat transfer.
A three-dimensional optimal solution emerges from bifurcation on the two-dimensional solution branch at higher $Pe$.
At $Pe\gtrsim10^{3}$, the optimised velocity fields consist of convection cells with hierarchical self-similar vortical structures, and the temperature fields exhibit a logarithmic mean profile near the walls.
\end{abstract}

\begin{keywords}
variational methods, mixing enhancement, B\'enard convection
\end{keywords}
 
\section{Introduction}
What is a flow optimising heat transfer?
We have explored an answer to this naive question.
For buoyancy-driven convection, i.e. Rayleigh--B\'enard convection, the maximal heat transfer has been discussed for more than half a century \citep{Malkus1954,Howard1963,Busse1969}.
\cite{Kraichnan1962} has predicted the asymptotic scaling of the Nusselt number $Nu$ with the Rayleigh number $Ra$ as $Nu\sim Ra^{1/2}$ with logarithmic correction for very high $Ra$.
In 1990's, a new variational approach called `the background method' was invented by \cite{Doering1992}, and the method has triggered remarkable advancements in the theoretical estimate of the upper bound on the Nusselt number $Nu$ \citep{Doering1996,Kerswell2001,Otero2002,Plasting2003,Doering2006,Whitehead2011,Whitehead2012}.
In these theoretical works, rigorous upper bounds, e.g. $Nu-1\le0.02634Ra^{1/2}$ \citep{Plasting2003}, have been derived at $Ra\gg 1$.
The `ultimate' law $Nu\sim Ra^{1/2}$ corresponds to the Taylor law of energy dissipation in high-Reynolds-number turbulence.
It has not been demonstrated as yet what flow structure achieves the ultimate scaling $Nu\sim Ra^{1/2}$.
Recently, meanwhile, \cite{Hassanzadeh2014} have numerically maximised a wall heat flux within a two-dimensional velocity field bounded by two parallel plates with a constant temperature difference.
They formulated a variational problem to find a velocity field maximising heat transfer under the constraint of fixed total enstrophy, and found optimal states consisting of an array of large-scale convection rolls for free-slip boundary conditions.
The maximal scaling is represented by $Nu\sim Ra^{5/12}$, corresponding to the rigid upper bound derived by the background method for free-slip conditions \citep{Whitehead2011,Whitehead2012}.
For no-slip conditions, on the other hand, the velocity fields numerically optimised within a two-dimensional field also exhibit large-scale circulation rolls, and the found scaling is $Nu\sim Ra^{0.37}$ \citep{Souza2016}.
Such scalings observed in the two-dimensional optimal states are quite distinct from the ultimate scaling $Nu\sim Ra^{1/2}$.

In this paper, we consider the variational problem first examined by \cite{Hassanzadeh2014} for free-slip conditions and then by \cite{Souza2016} for no-slip conditions; however, we extend a search range of optimal states to a three-dimensional velocity field.
We report three-dimensional optimal states capable of achieving the ultimate scaling, and discuss the optimised flow structures.
In order to satisfy the Navier--Stokes equation the optimised divergence-free vector field needs external body force which is distinct from buoyancy, but hereafter we refer to it as a `velocity' field.

\begin{figure}
\centering
\vspace{1em}
	\begin{minipage}{.35\linewidth}
	\includegraphics[clip,width=\linewidth]{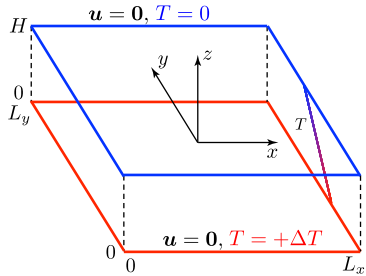}
	\end{minipage}
\caption{Configuration of the velocity and temperature fields.
\label{fig1}}
\end{figure}

\section{Formulation}
Let us consider heat transfer in a three-dimensional, time-independent and incompressible velocity field between two parallel plates,
$\mbox{\boldmath$u$}'(x',y',z')=u'\mbox{\boldmath$e$}_{x}+v'\mbox{\boldmath$e$}_{y}+w'\mbox{\boldmath$e$}_{z}$, satisfying the continuity equation
\begin{equation}
\label{eq2-1}
\nabla'\cdot\mbox{\boldmath$u$}'=0,
\end{equation}
where a prime $(\cdot)'$ represents a dimensional quantity, and $\mbox{\boldmath$e$}_{x}$ and $\mbox{\boldmath$e$}_{y}$ are mutually orthogonal unit vectors in the wall-parallel directions while $\mbox{\boldmath$e$}_{z}$ is a unit vector in the wall-normal direction.
The configuration of the velocity and temperature fields is shown in figure \ref{fig1}.
The two parallel plates are positioned at $z'=0$ and $z'=H$, and the domain of the flow is periodic in the $x$- and $y$-directions with periods, $L_{x}'$ and $L_{y}'$.
The upper (or lower) wall surface is held at lower (or higher) constant temperature $T'=0$ (or $T'=\Delta T>0$).
We suppose that the temperature field $T'(x',y',z')$ is determined as a solution to an advection-diffusion equation
\begin{equation}
\label{eq2-2}
(\mbox{\boldmath$u$}'\cdot\nabla') T'=\kappa\nabla'^{2}T',
\end{equation}
supplemented by the boundary conditions
\begin{eqnarray}
\label{eq2-3}
\mbox{\boldmath$u$}'(z'=0)=\mbox{\boldmath$u$}'(z'=H)=\mbox{\boldmath$0$};\hspace{1em}T'(z'=0)=\Delta T,\hspace{1em}T'(z'=H)=0,
\end{eqnarray}
where $\kappa$ denotes a thermal diffusivity.
The strength of the velocity field is measured by the P\'eclet number $Pe$ defined, in terms of the total enstrophy (or the averaged  square of velocity gradient tensor), as
\begin{equation}
\label{eq2-4}
Pe=\frac{{\left< |\mbox{\boldmath$\omega$}'|^{2} \right>}^{\frac{1}{2}} H^{2}}{\kappa}=\frac{{\left< |\nabla'\mbox{\boldmath$u$}'|^{2} \right>}^{\frac{1}{2}} H^{2}}{\kappa},
\end{equation}
where $\mbox{\boldmath$\omega$}'=\nabla'\times\mbox{\boldmath$u$}'$,
$|\nabla'\mbox{\boldmath$u$}'|^{2}=\nabla'\mbox{\boldmath$u$}':\nabla'\mbox{\boldmath$u$}'$ and $\left< \cdot \right>$ is a volume average.
The wall-normal convective heat transfer is characterized by the Nusselt number defined as the ratio of the convective heat flux to the conductive one,
\begin{equation}
\label{eq2-5}
Nu=1+\frac{\left< w'T' \right>}{\kappa \Delta T/H}.
\end{equation}
In this study, we explore a three-dimensional velocity field maximising $Nu$ for fixed $Pe$.
The constrained optimisation is relevant to the maximisation of the objective functional
\begin{eqnarray}
\label{eq2-6}
\displaystyle
\mathcal{F}=\Bigl< w\theta-\theta^{*}(\mbox{\boldmath$x$})\left[(\mbox{\boldmath$u$}\cdot\nabla)\theta-\nabla^{2}\theta-w \right]+p^{*}(\mbox{\boldmath$x$})\left( \nabla\cdot\mbox{\boldmath$u$} \right)+\frac{\mu}{2}\left( Pe^{2}-|\nabla\mbox{\boldmath$u$}|^{2}\right) \Bigr>
\end{eqnarray}
\citep[see][]{Hassanzadeh2014}, where $p^{*}(\mbox{\boldmath$x$})$, $\theta^{*}(\mbox{\boldmath$x$})$ and $\mu$ are Lagrange multipliers.
The variables in (\ref{eq2-6}) have been non-dimensionalised as
\begin{eqnarray}
\label{eq2-7}
\mbox{\boldmath$x$}=\frac{\mbox{\boldmath$x$}'}{H},\hspace{1em}\theta=\frac{\theta'}{\Delta T},\hspace{1em}\mbox{\boldmath$u$}=\frac{\mbox{\boldmath$u$}'}{\kappa/H},\hspace{1em}p^{*}=\frac{{p^{*}}'}{\rho\kappa^2/H^2},\hspace{1em}\theta^{*}=\frac{{\theta^{*}}'}{\Delta T},
\end{eqnarray}
where $\rho$ is the mass density of the fluid and $\theta=T-(1-z)$ is a temperature fluctuation about a conductive state.
Stationary points of $\mathcal{F}$ are determined by the Euler--Lagrange equations
\begin{eqnarray}
\label{eq2-8}
\displaystyle
\frac{\delta \mathcal{F}}{\delta \mbox{\boldmath$u$}}&\equiv&
 -\nabla p^{*}
+\theta\nabla\theta^{*}
+\mu\nabla^{2}\mbox{\boldmath$u$}
+(\theta+\theta^{*})\mbox{\boldmath$e$}_{z}=\mbox{\boldmath$0$},\\
\label{eq2-9}
\displaystyle
\frac{\delta \mathcal{F}}{\delta \theta}&\equiv&
(\mbox{\boldmath$u$}\cdot\nabla)\theta^{*}
+\nabla^{2}\theta^{*}
+w=0, \\
\label{eq2-10}
\displaystyle
\frac{\delta \mathcal{F}}{\delta \theta^{*}}&\equiv&
-(\mbox{\boldmath$u$}\cdot\nabla)\theta
+\nabla^{2}\theta
+w=0, \\
\label{eq2-11}
\displaystyle
\frac{\delta \mathcal{F}}{\delta p^{*}}&\equiv&
\nabla\cdot\mbox{\boldmath$u$}=0,\\
\label{eq2-12}
\displaystyle
\frac{\partial \mathcal{F}}{\partial \mu}&\equiv&
\frac{1}{2}{\left< Pe^{2}-|\nabla\mbox{\boldmath$u$}|^{2} \right>}=0.
\end{eqnarray}
 
\section{Numerical optimisation}
Solutions to equations (\ref{eq2-8})--(\ref{eq2-11}) depend only on $\mu$ for fixed periods $(L_{x},L_{y})$.
For given $\mu$, the solutions correspond to stationary points of the alternative functional
\begin{eqnarray}
\label{eq3-1}
\mathcal{G}&=&\Bigl< w\theta-\frac{\mu}{2}|\nabla\mbox{\boldmath$u$}|^{2}-\theta^{*}(\mbox{\boldmath$x$})\left[(\mbox{\boldmath$u$}\cdot\nabla)\theta-\nabla^{2}\theta-w \right]+p^{*}(\mbox{\boldmath$x$})\left( \nabla\cdot\mbox{\boldmath$u$} \right) \Bigr>.
\end{eqnarray}
This is because $\mathcal{G}=\mathcal{F}-(\mu/2)Pe^{2}$ and thus the Euler--Lagrange equations for $\mathcal{G}$ are also given by (\ref{eq2-8})--(\ref{eq2-11}).
In our previous work on a different functional in a different configuration \citep{Motoki2018}, we have developed a numerical approach to find local maxima of a functional kindred to $\mathcal{G}$ by a combination of the steepest ascent method and the Newton--Krylov method.
Using the same procedures, we obtain an optimal state ($\mbox{\boldmath$u$}_{\rm opt},\theta_{\rm opt},\theta^{*}_{\rm opt},p^{*}_{\rm opt}$) maximising $\mathcal{G}$ for given $\mu$.
Since $\mathcal{F}$ has the gradients common to $\mathcal{G}$, the optimal state gives the maximum of $\mathcal{F}$ at $Pe={\left< |\nabla\mbox{\boldmath$u$}_{\rm opt}|^{2} \right>}^{1/2}$.
Thus the optimal states of $\mathcal{F}$ can be obtained without fixing $Pe$ in the process of the optimisation.
Maximal points for a specific value of $Pe$ (say, $Pe_{0}$) are calculated by updating $\mu$ as
\begin{eqnarray}
\label{eq3-2}
\mu_{\rm new}=\mu+\epsilon ({\left< |\nabla\mbox{\boldmath$u$}_{\rm opt}|^{2} \right>}-Pe_{0}^{2}),
\end{eqnarray}
taking account of the fact that the decrease (or increase) in $\mu$ corresponds to the increase (or decrease) in $Pe$, where $\epsilon$ is a small positive constant. 
Equations (\ref{eq2-8})--(\ref{eq2-11}) are discretised employing the spectral Galerkin method based on
Fourier--Chebyshev expansions \citep[for more details, see section 3 and appendix A in ][]{Motoki2018}.

In this paper, we present the optimal states in the square wall-parallel domain of $(L_{x},L_{y},L_{z})=(\pi/2,\pi/2,1)$.
The numerical computations are carried out on $64^3$ grid points for $Pe\le5000$ and $128^3$ for $Pe>5000$.

\section{Ultimate scaling}
\begin{figure}
\centering
\vspace{1em}
	\begin{minipage}{.7\linewidth}
	(\textit{a})\\
	\includegraphics[clip,width=\linewidth]{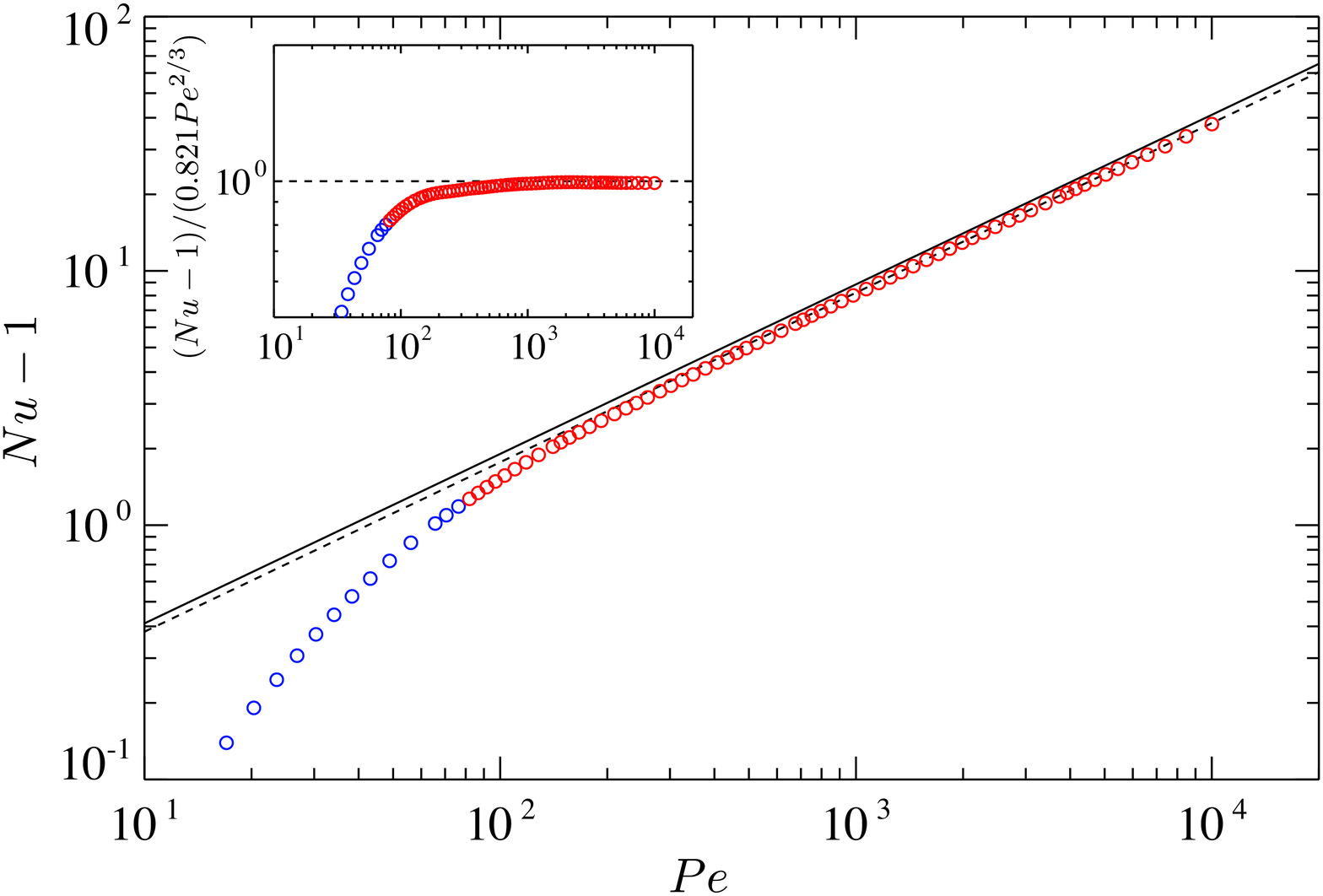}
	\end{minipage}
	
	\begin{minipage}{.4\linewidth}
	(\textit{b})\\
	\includegraphics[clip,width=\linewidth]{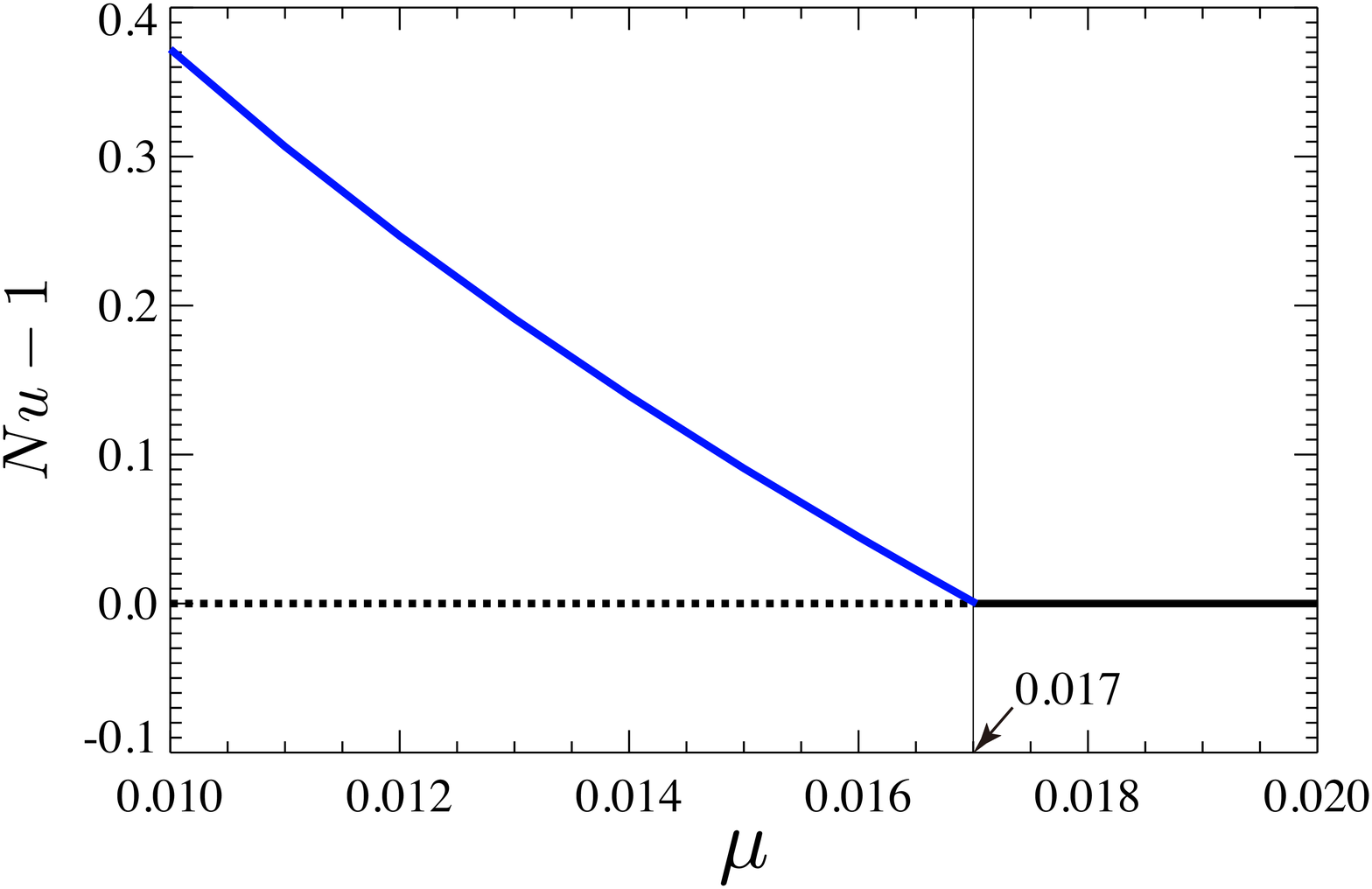}
	\end{minipage}
	\begin{minipage}{.4\linewidth}
	(\textit{c})\\
	\includegraphics[clip,width=\linewidth]{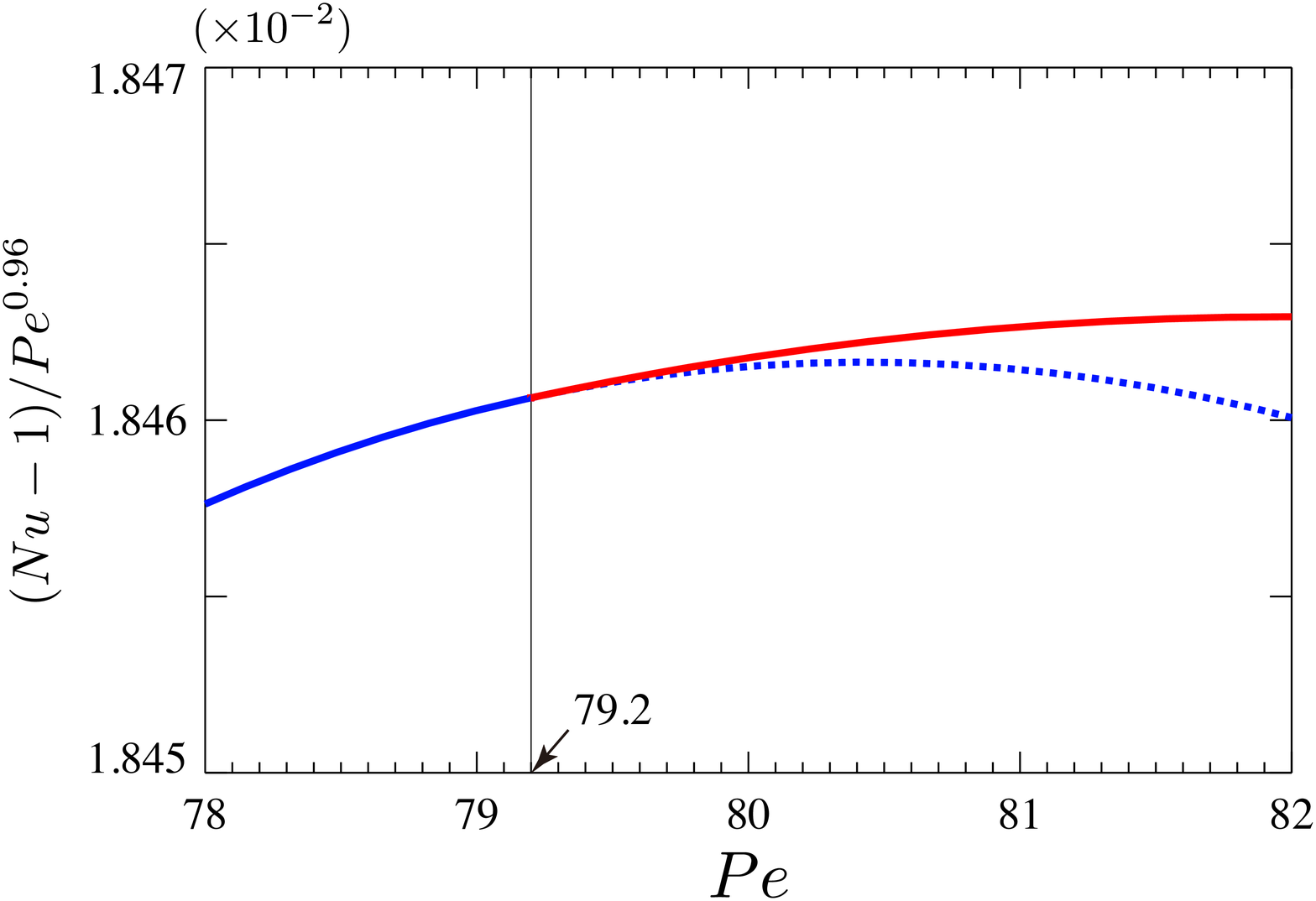}
	\end{minipage}
\caption{Nusselt number $Nu$ as a function of P\'eclet number $Pe$ in the optimal states.
The blue and red circles denote two-dimensional and three-dimensional optimal states, respectively.
The dashed line indicates the power fit $Nu-1=0.0821Pe^{2/3}$ determined in the range $5\times10^{3}<Pe<10^{4}$.
The solid line represents the scaling $Nu-1=0.0885Pe^{2/3}$ evaluated from the rigorous upper bound $Nu-1=0.02634Ra^{1/2}$ \citep{Plasting2003} assuming the identity $Pe^{2}=Ra(Nu-1)$ \citep{Hassanzadeh2014}.
The inset shows the compensated $Nu$.
({\it b,c}) $Nu$ as a function of ({\it b}) much larger $\mu$ (much smaller $Pe$) and ({\it c}) larger $\mu$ (smaller $Pe$).
The blue and red curves respectively show the two-dimensional and three-dimensional solutions, and the black one is a conductive solution.
The solid (or dashed) curve denotes an optimal (or saddle) solution.
\label{fig2}}
\end{figure}

Figure \ref{fig2}({\it a}) shows the maximal $Nu$ as a function of $Pe$.
At large $Pe$ ($>10^3$), we observe the scaling of $Nu$ with $Pe$, $Nu\sim Pe^{2/3}$.
The scaling $Nu\sim Pe^{2/3}$ corresponds to the ultimate scaling $Nu\sim Ra^{1/2}$ in the Rayleigh--B\'enard problem, provided that the total energy budget is given by the Boussinesq equation, that is $Pe^{2}=Ra(Nu-1)$ \citep{Hassanzadeh2014}, where $Ra=g\beta\Delta TH^{3}/(\nu\kappa)$ is the Rayleigh number, $g$ and $\beta$ being the acceleration due to gravity and the thermal expansion coefficient of the fluid, respectively.
The thick solid line indicates the rigorous upper bound derived by using the background method \citep{Plasting2003}.
The obtained maximal scaling is close to the upper bound, and the prefactor is about 7.2\% less than that of the bound.

Choosing the reference velocity as $U=(g\beta\Delta T H)^{1/2}$, we have the scaling with respect to the energy dissipation as
\begin{eqnarray}
\label{eq4-1}
\frac{\nu{\left< |\nabla'\mbox{\boldmath$u$}'|^{2} \right>}}{U^{3}/H}\sim Pr^{-1/2},
\end{eqnarray}
where $\nu$ and $Pr$ are the kinematic viscosity and the Prandtl number, respectively.
Thus the scaling $Nu\sim Pe^{2/3}$ means that the energy dissipation normalised by $U^{3}/H$ is independent of the 
Reynolds number, in accord with the Taylor's scaling view for turbulent energy dissipation.
For homogeneous turbulent convection without thermal and velocity boundary layers, e.g. in three-dimensional periodic boundary box with the vertical mean temperature gradient \citep{Lohse2003}, the ultimate scaling $Nu\sim Ra^{1/2}$ has been observed.
Although the Taylor dissipation law also does not hold in turbulent shear flows over a smooth wall surface, the Reynolds-number-independent skin-friction coefficient can be observed in high-Reynolds-number rough-wall turbulence, implying the emergence of the Taylor law.
However, it has still been an open question whether or not the ultimate scaling can be found in high-$Ra$ convective turbulence between two parallel plates with surface roughness \citep{Roche2001,Zhu2017}.

\begin{figure}
\centering
\vspace{1em}
        \begin{minipage}{.3\linewidth}
	(\textit{a})\\
	\includegraphics[clip,width=\linewidth]{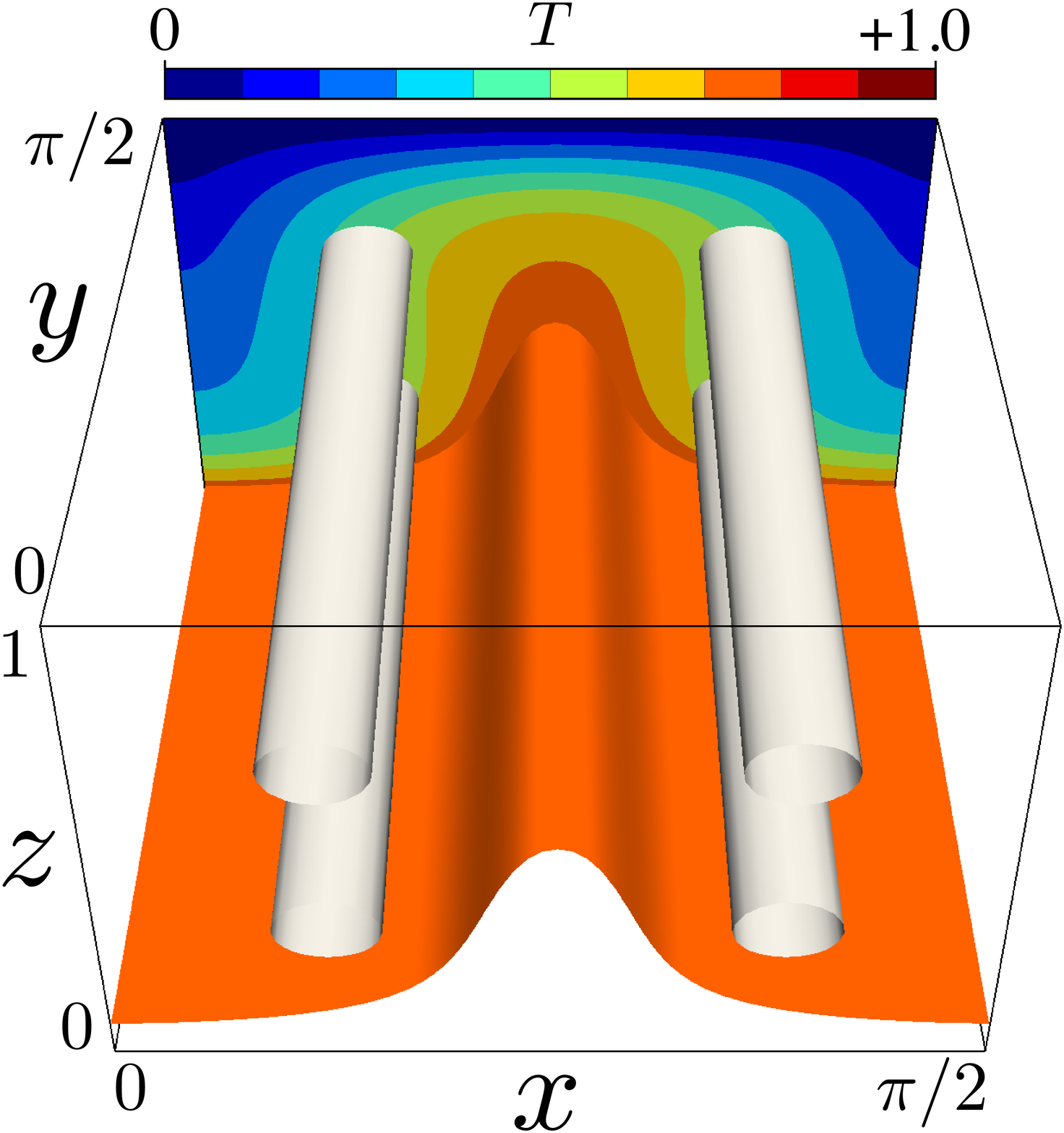}
	\end{minipage}
	\hspace{2em}
	\begin{minipage}{.3\linewidth}
	(\textit{b})\\
	\includegraphics[clip,width=\linewidth]{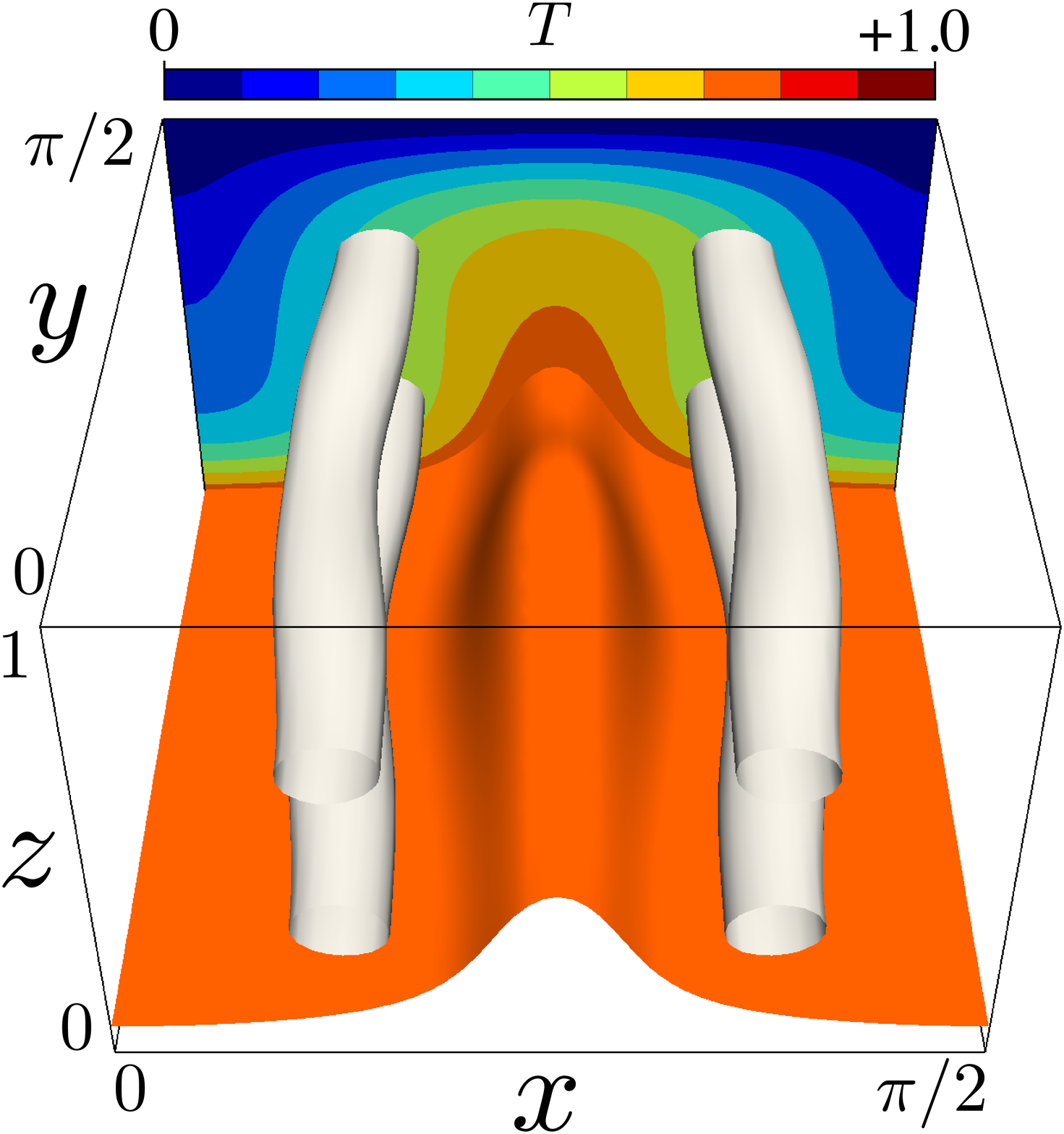}
	\end{minipage}

        \begin{minipage}{.3\linewidth}
	(\textit{c})\\
	\includegraphics[clip,width=\linewidth]{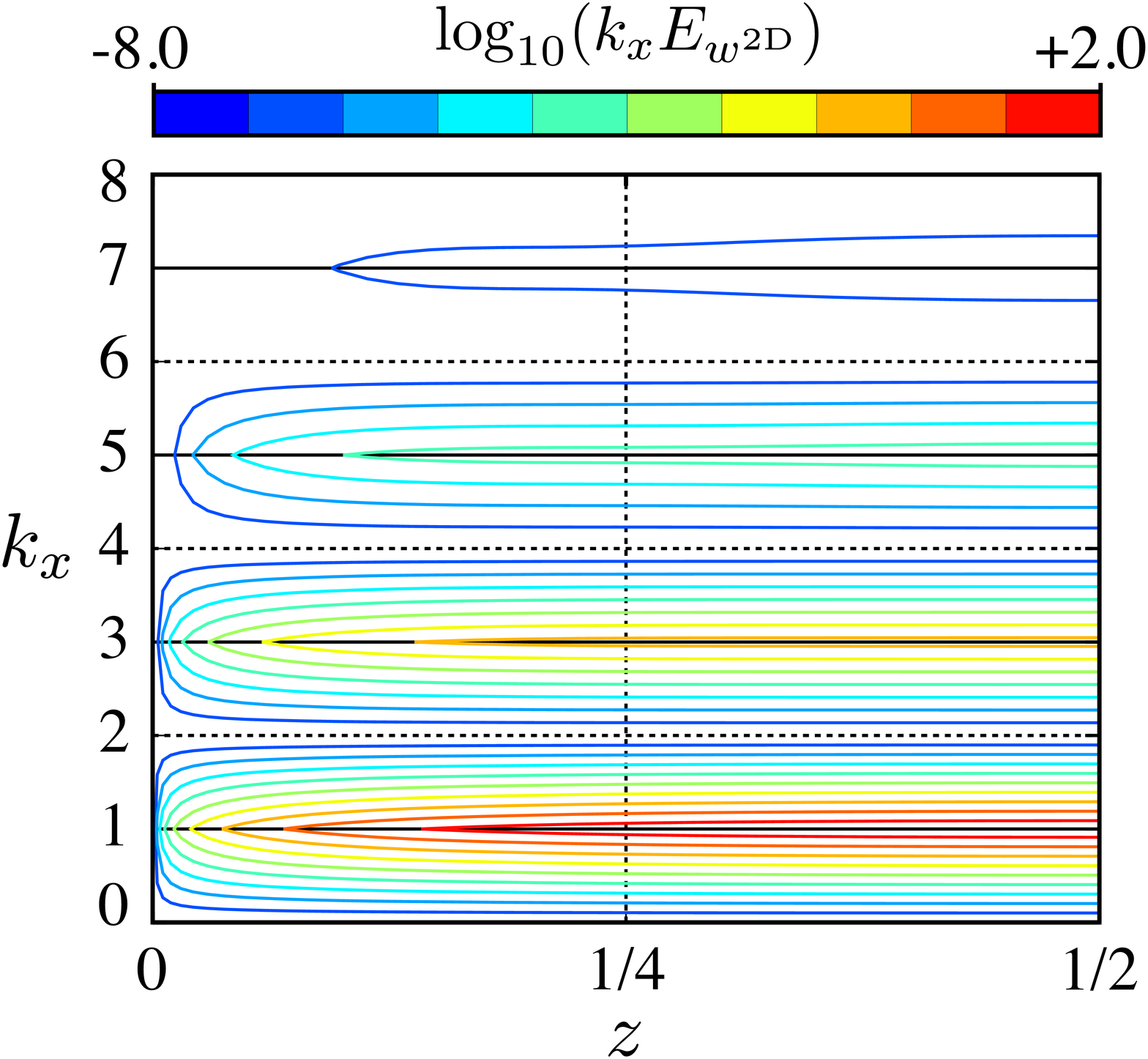}
	\end{minipage}
	\hspace{2em}
	\begin{minipage}{.3\linewidth}
	(\textit{d})\\
	\includegraphics[clip,width=\linewidth]{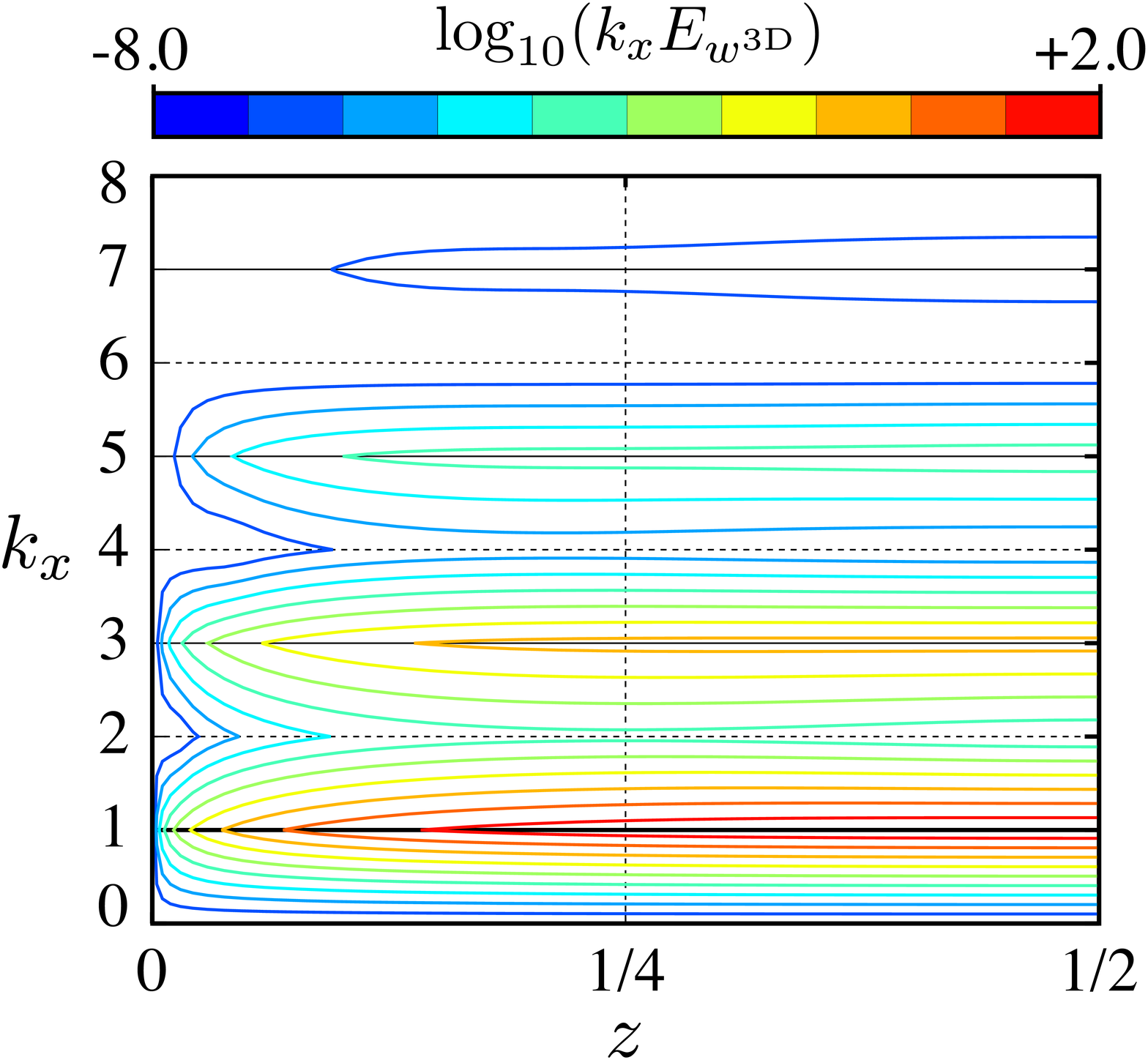}
	\end{minipage}

        \begin{minipage}{.3\linewidth}
	(\textit{e})\\
	\includegraphics[clip,width=\linewidth]{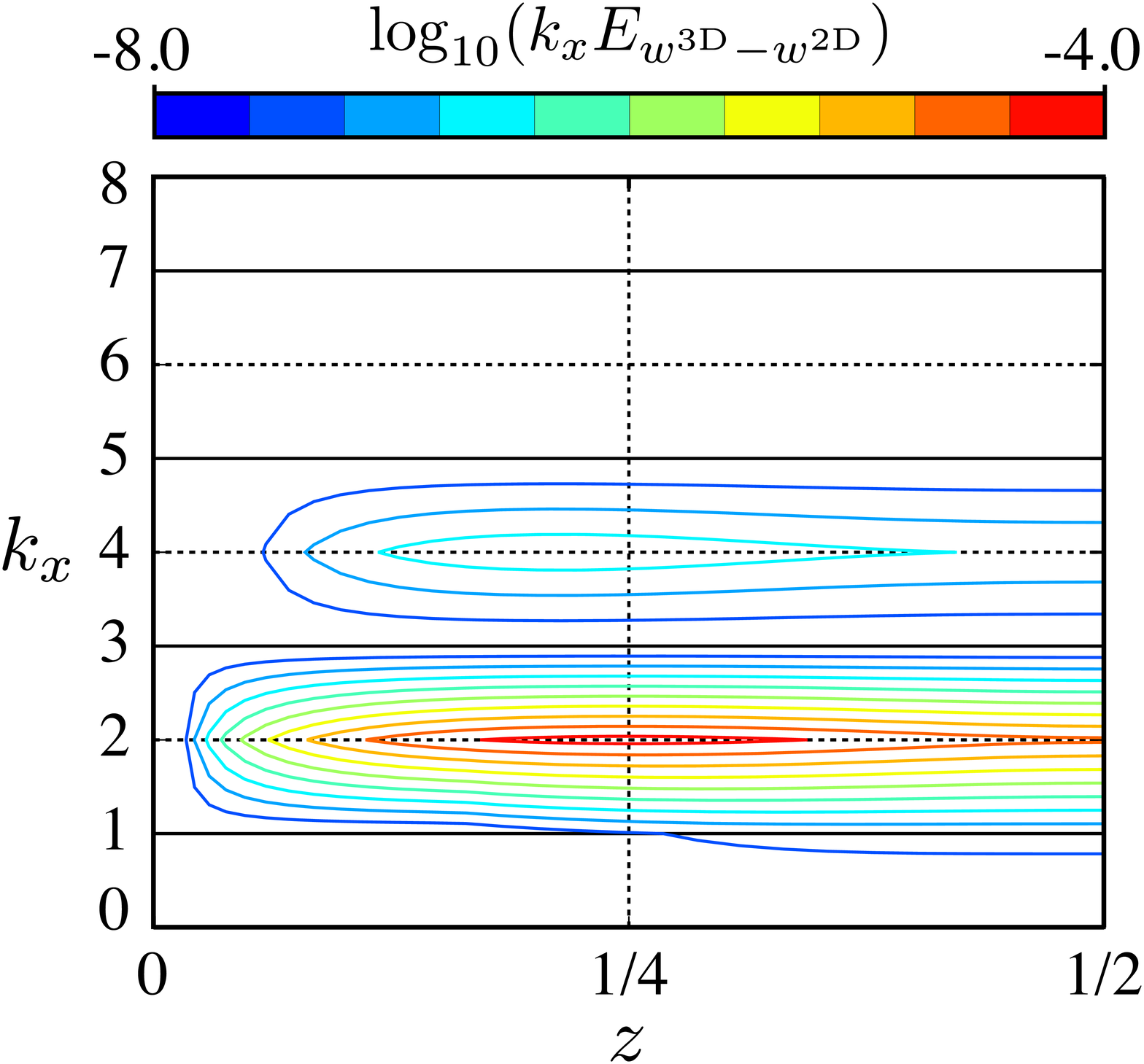}
	\end{minipage}
	\hspace{2em}
	\begin{minipage}{.3\linewidth}
	(\textit{f})\\
	\includegraphics[clip,width=\linewidth]{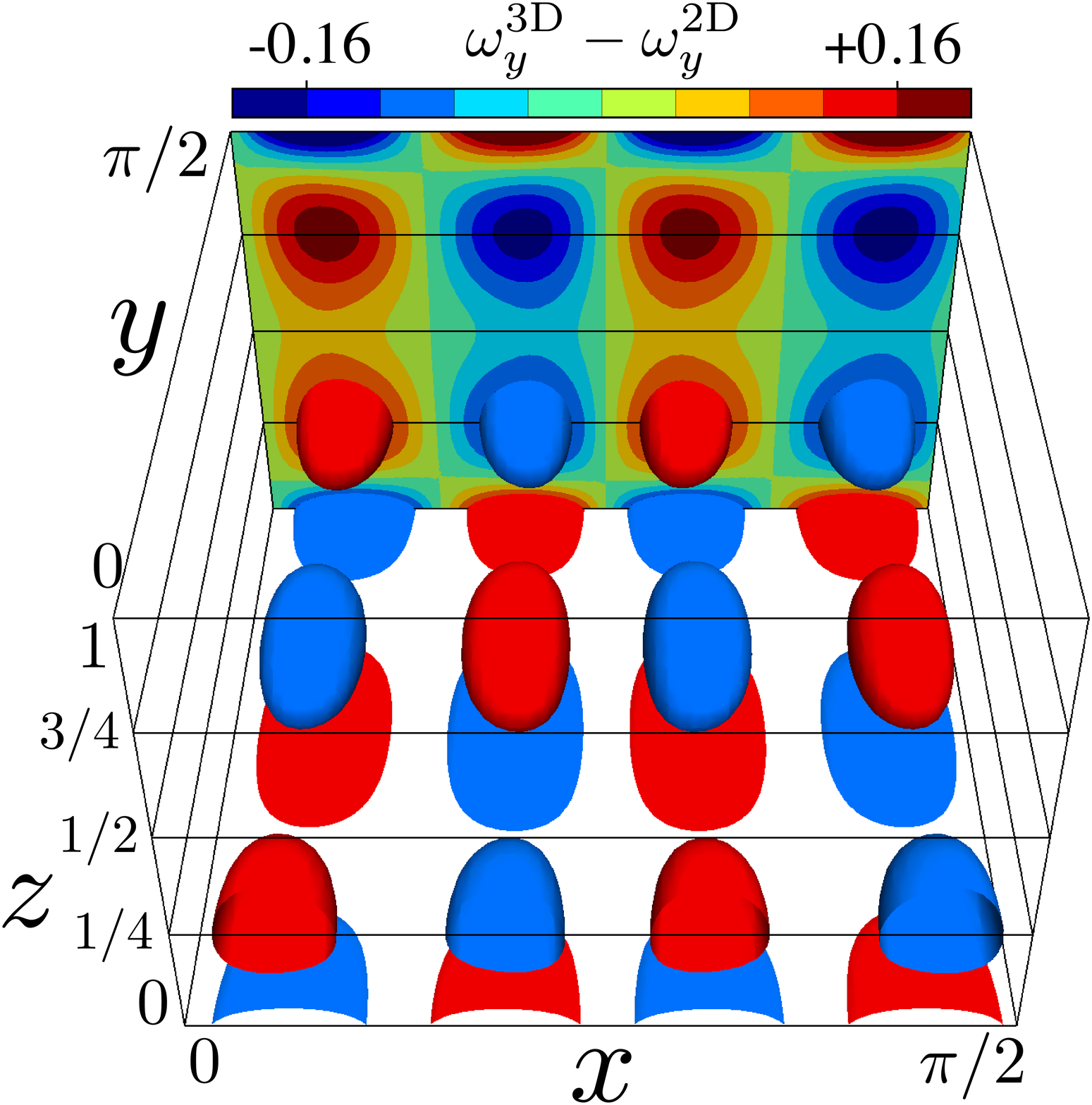}
	\end{minipage}
\caption{({\it a}) Two-dimensional saddle and ({\it b}) three-dimensional optimal solutions at $Pe=80.0$.
The orange objects show the isosurfaces of the temperature, $T=0.75$, and the white tube-like objects are the vortex structures visualised by the positive second invariant of the velocity gradient tensor, $Q=2560$.
The contours represent the temperature field in the plane $y=\pi/2$.
({\it c--e}) One-dimensional pre-multiplied energy spectra of the wall-normal velocity $w$, $k_{x}E_{w}$, at $Pe=79.2$.
The spectrum of ({\it c})
the two-dimensional solution $w^{\rm 2D}$ and ({\it d}) the three-dimensional solution $w^{\rm 3D}$; and of ({\it e}) their difference $w^{\rm 3D}-w^{\rm 2D}$.
The lateral axis denotes the distance to the wall $z$, and the longitudinal one is the wavenumber component $k_{x}$ in the $x$-direction.
({\it f}) Spatial distribution of the difference in the $y$-component of vorticity $\omega_{y}$ between the three- and two-dimensional solutions, $\omega_{y}^{\rm 3D}-\omega_{y}^{\rm 2D}$ at $Pe=79.2$.
The red/blue objects respectively show the isosurfaces of $\omega_{y}^{\rm \rm 3D}-\omega_{y}^{\rm \rm 2D}=\pm0.12$.
The contours represent $\omega_{y}^{\rm 3D}-\omega_{y}^{\rm 2D}$ in the plane $y=\pi/2$.
\label{fig3}}
\end{figure}

\section{Appearance of three-dimensional solution}
At large $\mu$ (small $Pe$), a two-dimensional array of convection rolls gives maximal heat transfer.
The solution arises from supercritical pitchfork bifurcation on a conductive solution at $\mu=1.703\times10^{-2}$ ($Pe\equiv0$) (figure \ref{fig2}{\it a}), and it satisfies the reflection symmetry
\begin{eqnarray}
\label{eq5-1}
[u,v,w,\theta](x,y,z)=[-u,v,w,\theta](-x,y,z)
\end{eqnarray}
and the shift-and-reflection symmetry
\begin{eqnarray}
\label{eq5-2}
[u,v,w,\theta](x,y,z)=[u,v,-w,-\theta](x+L_{x}/2,y,1-z)
\end{eqnarray}
(see figure \ref{fig3}{\it a}).
Figures \ref{fig3}({\it a,b}) visualise isosurfaces of the temperature field $T$ and of the second invariant of the velocity gradient tensor, $Q$.
As $\mu$ decreases further, the secondary pitchfork bifurcation occurs on the two-dimensional solution branch at $\mu=3.028\times10^{-3}$ ($Pe=79.2$) (figure \ref{fig2}{\it c}).
Subsequently, the two-dimensional solution becomes a saddle solution, and a three-dimensional optimal solution with the shift-and-reflection symmetry
\begin{eqnarray}
\label{eq5-2}
[u,v,w,\theta](x,y,z)=[u,v,-w,-\theta](x+L_{x}/2,y+L_{y}/2,1-z)
\end{eqnarray}
appears (see figure \ref{fig3}{\it b}).
Figures \ref{fig3}({\it c--e}) show the energy spectra of the wall-normal velocity $w$ at the onset of the three-dimensional solution as a function of the distance to the wall, $z$ and the $x$-component of the wavenumber vector, $k_{x}$.
In the two-dimensional solution shown in figure \ref{fig3}({\it c}), the wall-normal velocity $w$ consists of only odd-wavenumber components.
The spectral peak is located for $k_{x}=1$ at the midplane $z=1/2$, and it is relevant to the large-scale rolls.
The even-wavenumber components appear as a result of the bifurcation from the two-dimensional solution to the three-dimensional one (figure \ref{fig3}{\it d}).
The difference in the spectra between the two solutions at $Pe=79.2$ is shown in figure \ref{fig3}({\it e}).
The leading mode is at $k_{x}=2$, and the spectral component has a peak at $z=1/4$ (half the distance between one of the two walls and the midplane).
In figure \ref{fig3}({\it f}), the relevant structures are visualised by the difference in the $y$-component of vorticity, $\omega_{y}^{\rm 3D}-\omega_{y}^{\rm 2D}$.
The extracted structure is characterised in terms of a three-dimensional mode $(k_{x},k_{y})=(2,1)$, and exhibits an array of vortices arranged in a wall-parallel plane around $z=1/4$.
The onset of the smaller three-dimensional vortical structures near the walls brings about the bending of the original larger two-dimensional rolls and associated vortex tubes (figure \ref{fig3}{\it b}), enhancing heat transfer.

\section{Hierarchical self-similar structures}
\begin{figure}
\centering
\vspace{1em}
        \begin{minipage}{.4\linewidth}
	(\textit{a})\\
	\includegraphics[clip,width=\linewidth]{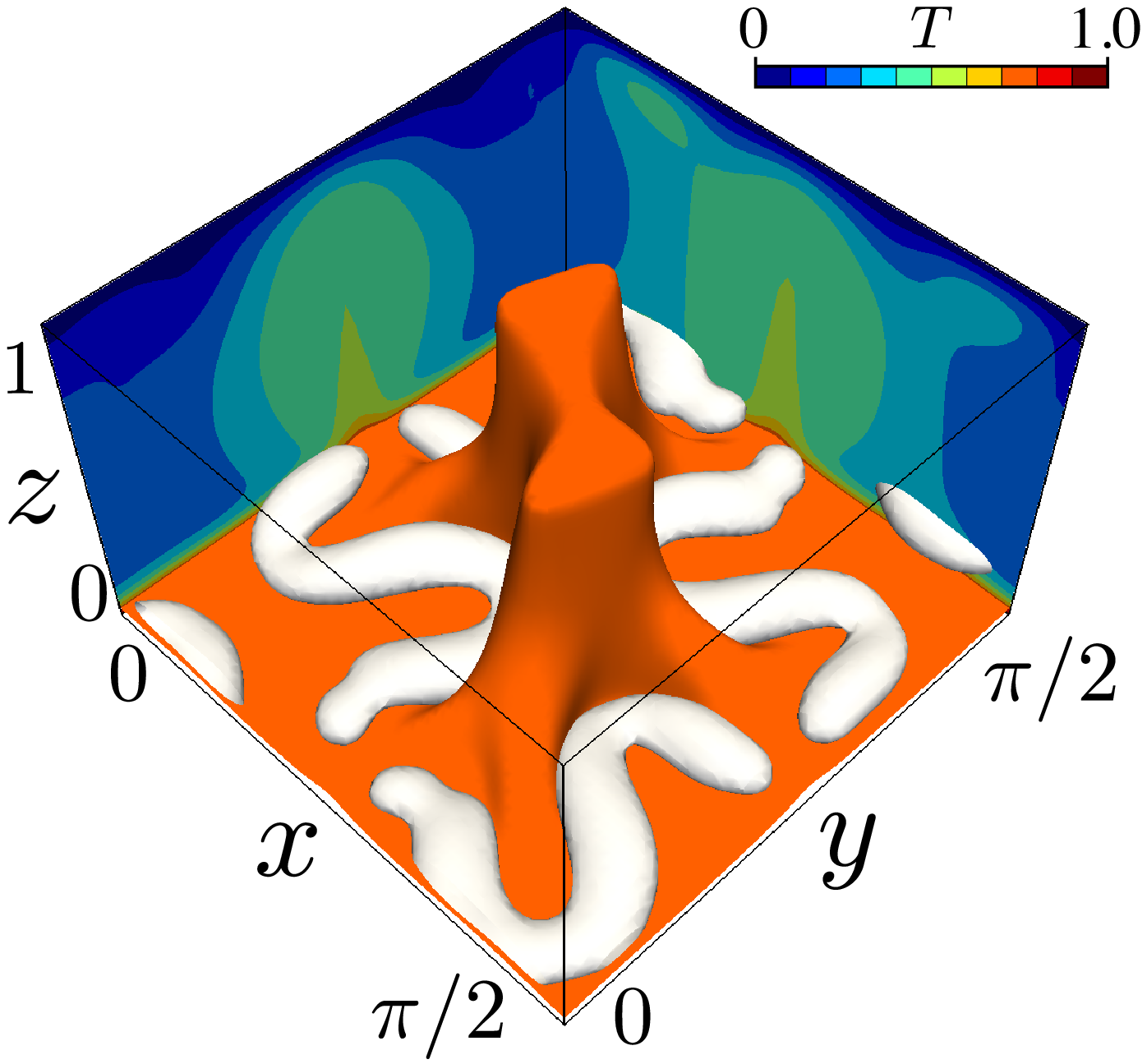}
	\end{minipage}
	\hspace{2em}
	\begin{minipage}{.4\linewidth}
	(\textit{b})\\
	\includegraphics[clip,width=\linewidth]{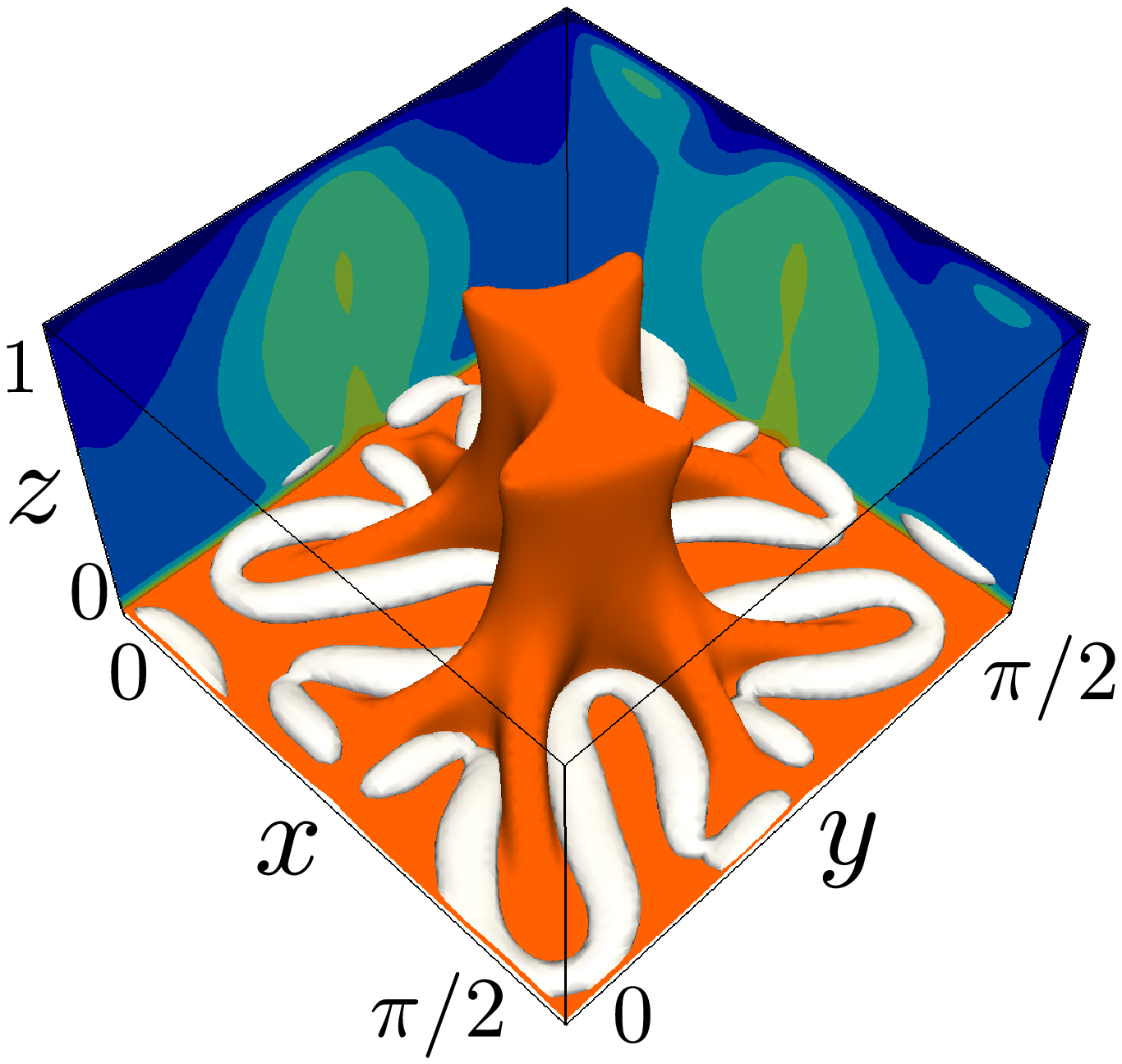}
	\end{minipage}
	
        \vspace{2em}
        \begin{minipage}{.4\linewidth}
	(\textit{c})\\
	\includegraphics[clip,width=\linewidth]{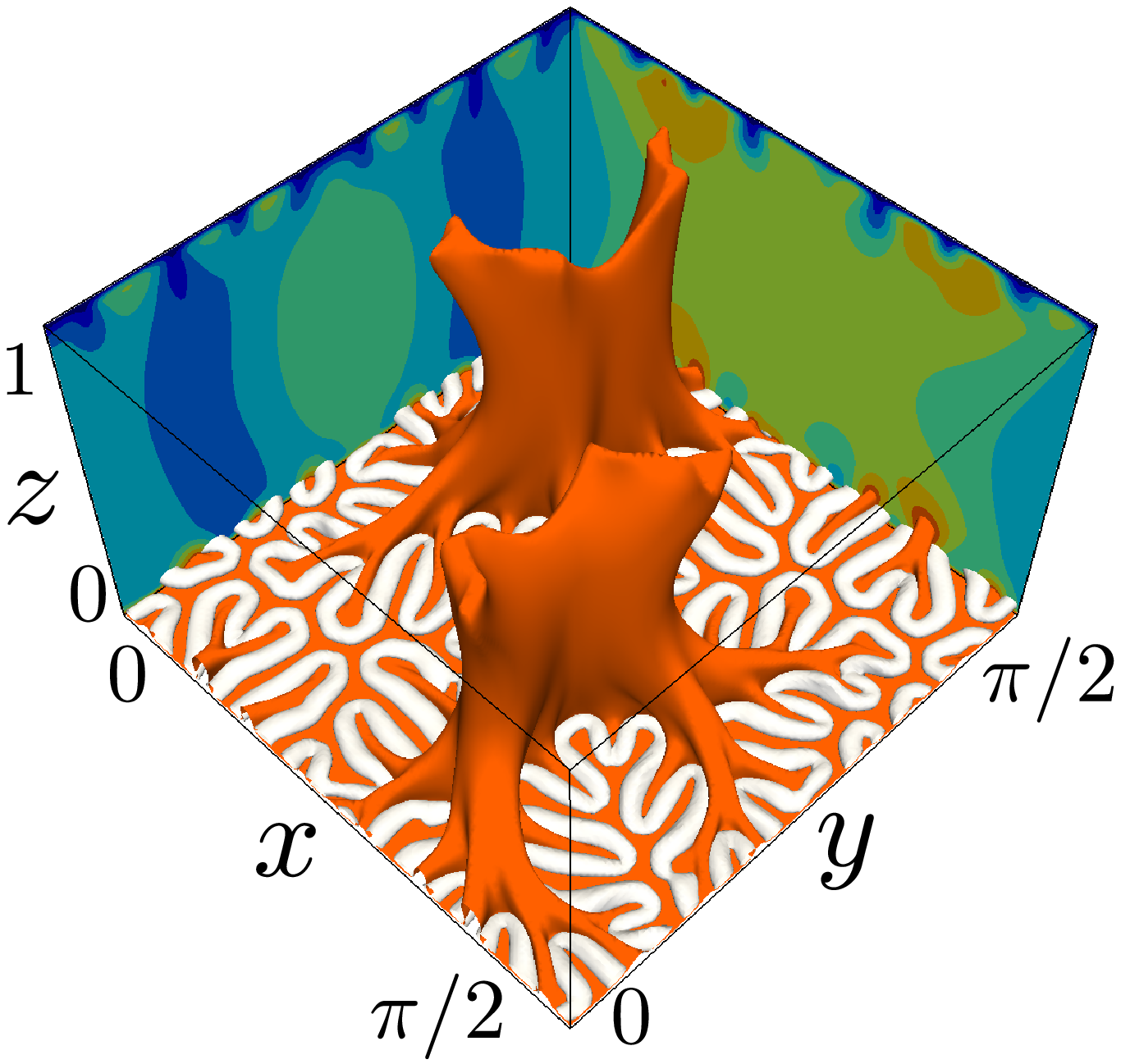}
	\end{minipage}
	\hspace{2em}
	\begin{minipage}{.4\linewidth}
	(\textit{d})\\
	\includegraphics[clip,width=\linewidth]{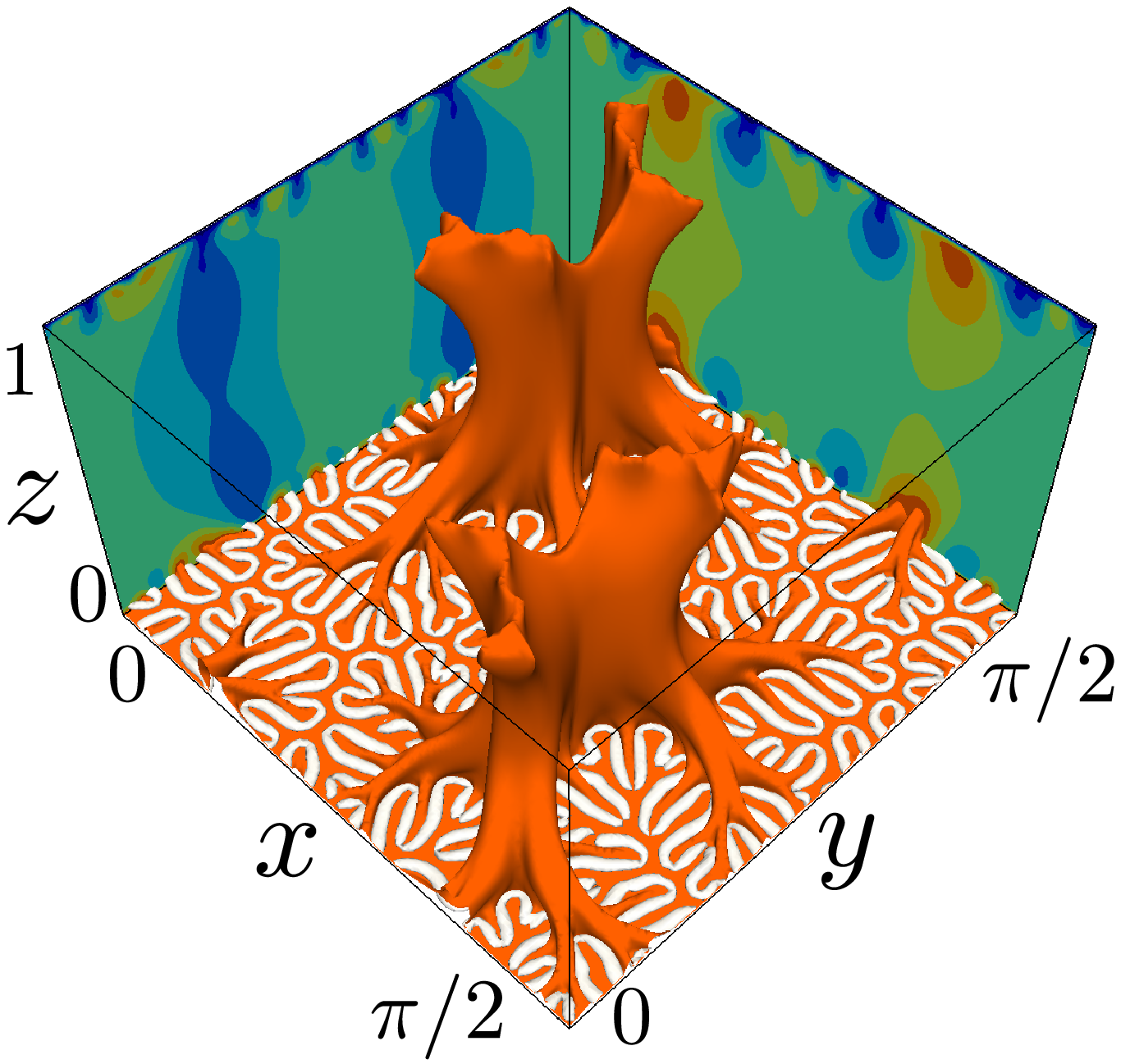}
	\end{minipage}
\caption{Optimal states at P\'eclet number ({\it a}) $Pe=508$, ({\it b}) $Pe=1006$, ({\it c}) $Pe=5041$ and ({\it d}) $Pe=10009$.
The orange objects show the isosurfaces of $T=0.75$.
The white tube-like structures are the isosurfaces of ({\it a}) $Q=8.0\times10^4$, ({\it b}) $Q=4.8\times10^4$, ({\it c}) $Q=1.6\times10^7$ and ({\it d}) $Q=1.6\times10^8$ (note that only those in the lower half of the domain are shown for visualisation of the near-wall structures).
The contours represent temperature field in the planes $x=\pi/2$ and $y=0$.
\label{fig4}}
\end{figure}

\begin{figure}
\centering
\vspace{1em}
        \begin{minipage}{.7\linewidth}
	\includegraphics[clip,width=\linewidth]{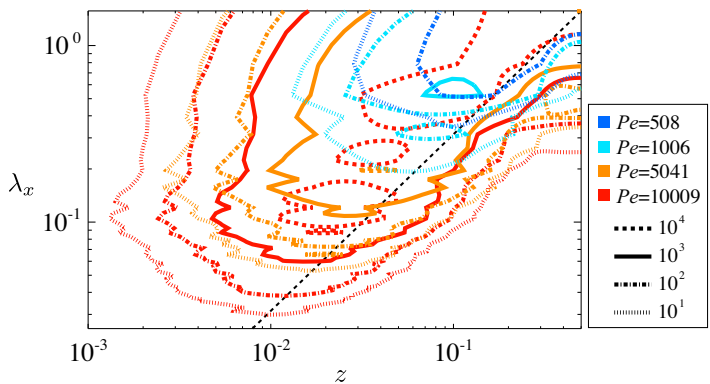}
	\end{minipage}
\caption{Energy spectra of the wall-normal velocity $w$, $k_{x}E_{w}$, as a function of the distance to the wall, $z$ and the wavelength in the $x$-direction, $\lambda_{x}$.
The dashed diagonal indicates $\lambda_{x}=L_{x}z$.
\label{fig5}}
\end{figure}

\begin{figure}
\centering
\vspace{1em}
        \begin{minipage}{.6\linewidth}
	(\textit{a})\\
	\includegraphics[clip,width=\linewidth]{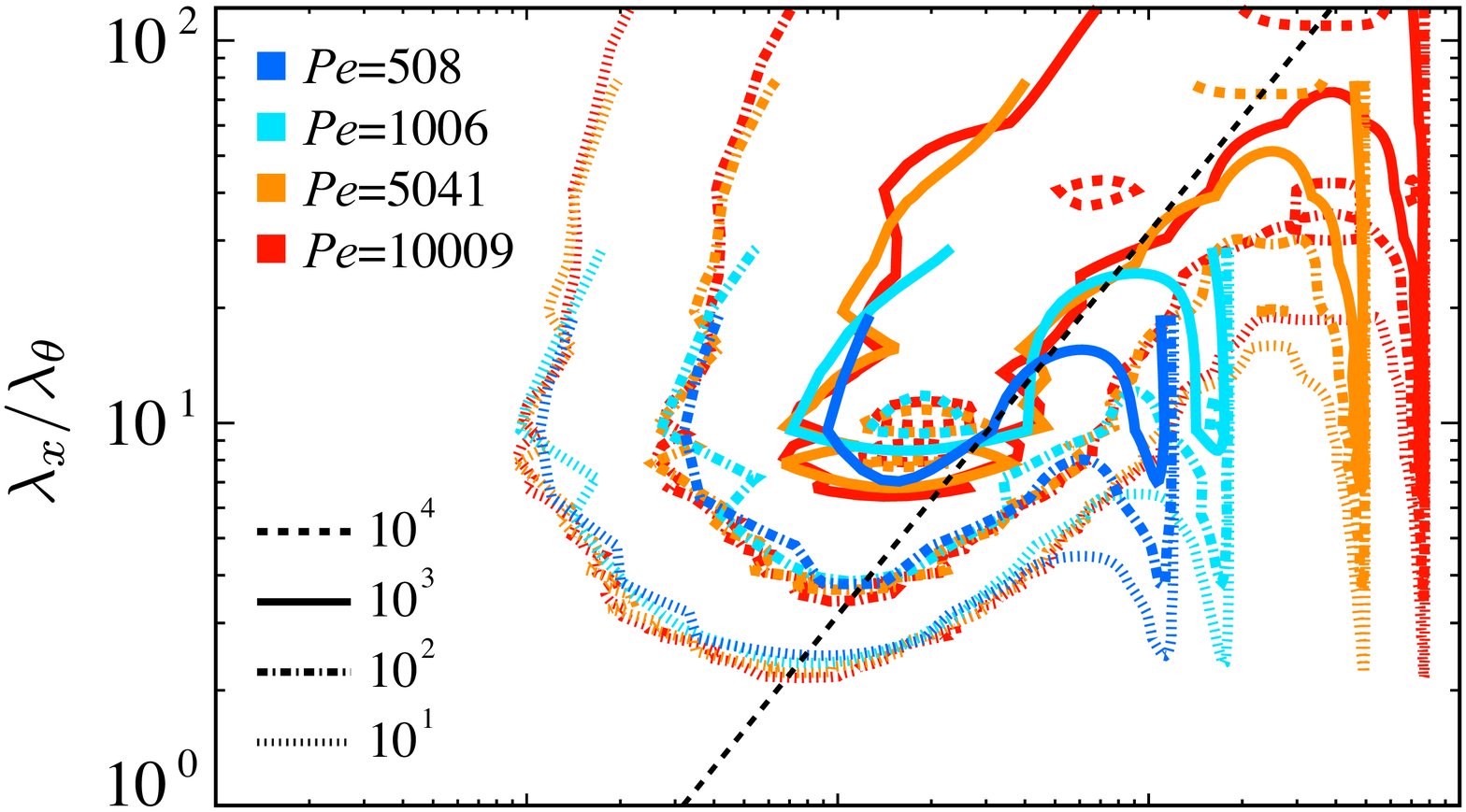}
	\end{minipage}

	\begin{minipage}{.6\linewidth}
	(\textit{b})\\
	\includegraphics[clip,width=\linewidth]{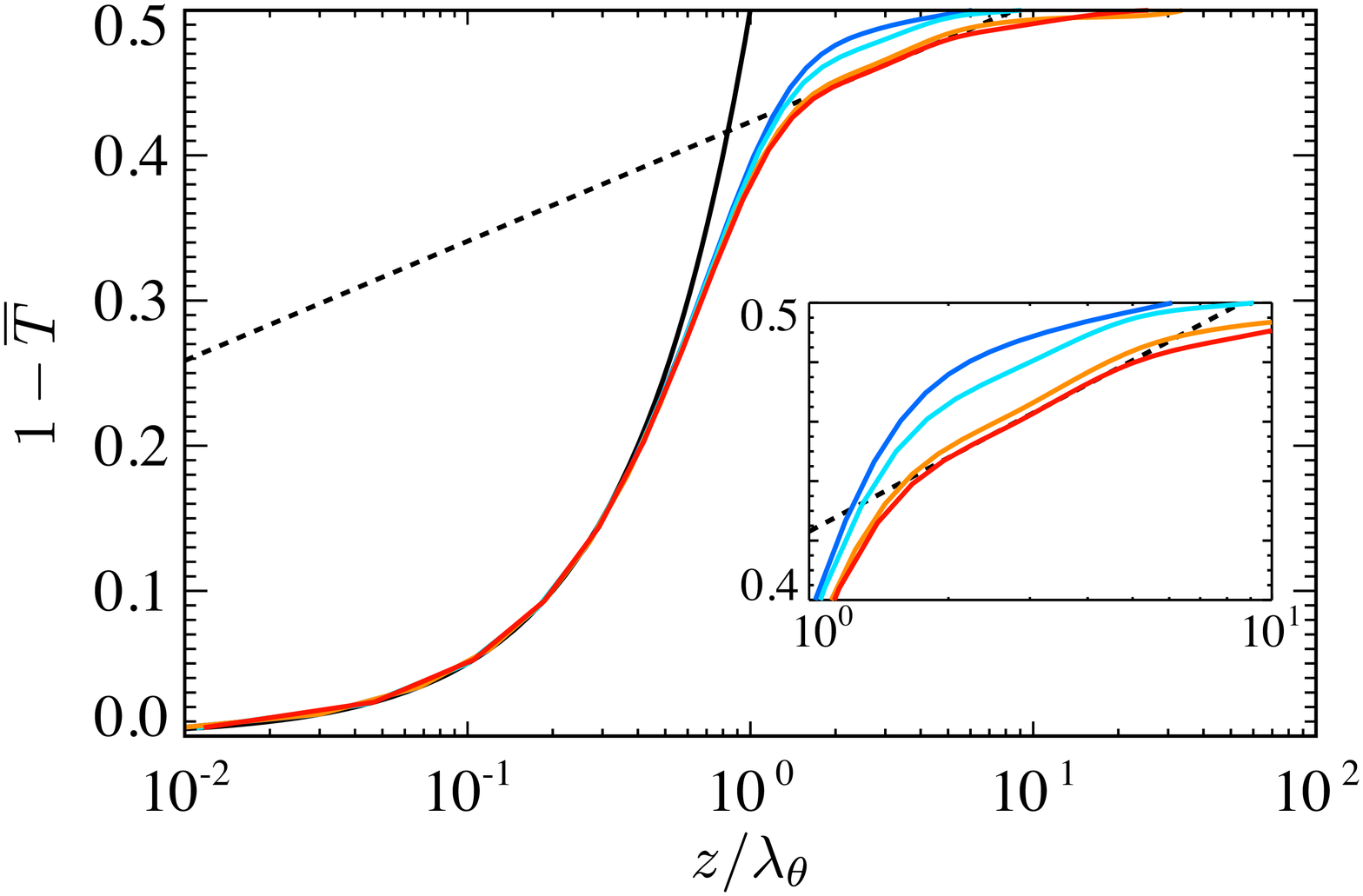}
	\end{minipage}
\caption{
({\it a}) Energy spectra $k_{x}E_{w}$ as a function of $z$ and $\lambda_x$.
The distance to the wall, $z$ and the wavelength in the $x$-direction, $\lambda_{x}$ are normalised by $\lambda_{\theta}=(2Nu)^{-1}$.
The dashed diagonal indicates $\lambda_{x}=L_{x}z$.
({\it b}) Mean temperature profile $\overline{T}$ as a function of $z/\lambda_{\theta}$.
The solid curve indicates $1-\overline{T}=z/\lambda_{\theta}$, and the dashed line represents the logarithmic fit $1-\overline{T}=0.0358\ln{(z/\lambda_{\theta})}+0.423$ determined in the range $2<z/\lambda_{\rm \theta}<4$ at $Pe=10009$.
\label{fig6}}
\end{figure}

Tree-like structure of isotherms is observed in the optimal states at small $\mu$ (large $Pe$), shown in figures \ref{fig4}.
The orange objects show isosurfaces of $T=0.75$, and a `trunk' of the `tree' represents a hot `plume' where the positive wall-normal velocity has been found to be dominant.
As $Pe$ increases, the tree `roots' grow deeper while maintaining the large-scale trunk.
The white objects show smallest-scale vortex structures visualised by the positive iso-surfaces of the second invariant of the velocity gradient tensor in the near-wall region of the lower half of the domain (similar vortical structures exist on the upper wall).
The smaller and stronger vortices appear closer to the walls with increasing the enstrophy, i.e. $Pe$.
The roots are seen to be generated as a consequence of upward fluid motion induced in between the roughly anti-parallel nearest segments of the winding tube-like vortices.
As seen in the bifurcation of the three-dimensional solution from the two-dimensional solution, the local folding of the larger vortices stems from the onset of the smaller vortical structures closer to the wall.
Figure \ref{fig5} shows the energy spectra of the wall-normal velocity $w$ as a function of the distance to the wall, $z$ and the wavelength in the $x$-direction, $\lambda_{x}=L_{x}/k_{x}$ relevant to the size of the vortical structures.
It can be seen that smaller-scale structures are generated closer to the wall as $Pe$ is increased.
At $Pe=10009$ several spectral peaks are observed along the `ridge' represented by the dashed diagonal $\lambda_{x}=L_{x}z$, implying that the optimal velocity fields possess hierarchical self-similarity.
As shown in figure \ref{fig6}({\it a}), the energy spectra scale with the conduction length $\lambda_{\theta}=(2Nu)^{-1}$ in the close vicinity of the wall.
The hierarchical structures exist down to $z/\lambda_{\theta}\approx1$, where the size of the structures is $\lambda_{x}\approx5Nu^{-1}$.
Since $Nu$ scales with $Pe^{2/3}$ at large $Pe$, the smallest scale is estimated as $\lambda_{x}\sim Pe^{-2/3}$ much smaller than the optimal aspect ratio, $L/H\sim Pe^{-0.371}$, in the two-dimensional field \citep{Souza2016}.
Figure \ref{fig6}({\it b}) shows the mean temperature profile $\overline{T}$ as a function of $z/\lambda_{\theta}$.
$1-\overline{T}=z/\lambda_{\theta}$ holds at $z/\lambda_{\theta}\ll1$, where the thermal conduction dominates over the convection.
As the distance to the wall, $z$ increases, the hierarchical vortex structures promote the convective heat transfer.
In the region $1\lesssim z/\lambda_{\theta}\lesssim10$, the logarithmic-like temperature profiles are observed at $Pe=1008,5041$ and $10009$.
The dashed line indicates the logarithmic fit $1-\overline{T}=0.0358\ln{(z/\lambda_{\theta})}+0.423$ determined in the range $2<z/\lambda_{\theta}<4$ at $Pe=10009$.
Recently, the logarithmic temperature profiles have also been observed numerically and experimentally in turbulent Rayleigh--B\'enard convection \citep{Ahlers2012,Ahlers2014}.
In the region far from the wall, $10\lambda_{\theta}\lesssim z\le1/2$, mixing by the large-scale convection cells is dominant, and thus the temperature profile is flattened.

\section{Summary and conclusions}
We have found the three-dimensional optimal states which lead to the scaling $Nu\sim Pe^{2/3}$ consistent with the ultimate scaling $Nu\sim Ra^{1/2}$ in
Rayleigh--B\'enard convection.
The optimal heat transfer is achieved by three-dimensional convection cells with smaller-scale vortices attached on the walls.
At large $Pe$, the optimal velocity field exhibits hierarchical self-similarity.
The large-scale cells mix up the temperature almost completely around the midplane between the two walls.
Near the walls, meanwhile, self-similar vortical structures locally enhance heat transfer, and yield logarithmic mean temperature distribution.
Our earlier optimisation for heat transfer in plane Couette flow \citep{Motoki2018} provided the optimal velocity fields in which we observed hierarchical structure consisting of a number of streamwise vortex tubes.
The logarithmic mean temperature profiles as well as the ultimate scaling $Nu\sim Ra^{1/2}$ were also found in the optimal fields.
It has recently been observed that the ultimate scaling $Nu\sim Ra^{1/2}$ can be achieved by some velocity field which is two-dimensional but exhibits hierarchical self-similarity \citep{Tobasco2017}.
These results suggest that self-similar hierarchy of a velocity field would be a necessary condition for the emergence of the ultimate scaling and logarithmic mean temperature profile between two-parallel no-slip plates.
The optimal state for heat transfer identified in this work should be closely relevant to convective turbulence, although external body force to be necessary for the optimal state to fulfill the Navier--Stokes equation is different from buoyant force in the Boussinesq equation.
Our preliminary study, in reality, demonstrates that by using homotopy from the body force to the buoyancy the optimal state can be continuously connected to a steady solution to the Boussinesq equation, which well represents the structure and statistics of convective turbulence.

\section*{Acknowledgements}
This work was partially supported by a Grant-in-Aid Scientific Research (grant nos. 25249014, 26630055) from the Japanese Society for Promotion of Science 665 (JSPS). S.M. is supported by JSPS Grant-in-Aid for JSPS Fellows Grant Number 666 16J00685.

\bibliographystyle{jfm}
\bibliography{ref}

\end{document}